\documentclass[aps,twocolumn,showpacs,floatfix]{revtex4}
\usepackage{graphics}
\usepackage{epsfig}
\usepackage{subfigure}
\usepackage{amsmath,amsfonts,amssymb,graphicx,times}
\begin{document}
\title{Memory-Based Snowdrift Game on Networks}
\author{Wen-Xu Wang$^{1,}$$^{2}$}
\author{Jie Ren$^{2}$}
\author{Guanrong Chen$^{1}$}\email{gchen@ee.cityu.edu.hk}
\author{Bing-Hong Wang$^{2}$}
\affiliation{$^{1}$Department of Electronic Engineering, City
University of Hong Kong, Hong Kong SAR, China\\
$^{2}$Department of Modern Physics, University of Science and
Technology of China, Hefei 230026, China}
\date{\today}

\begin{abstract}
We present a memory-based snowdrift game (MBSG) taking place on
networks. We found that, when a lattice is taken to be the
underlying structure, the transition of spatial patterns at some
critical values of the payoff parameter is observable for both $4$
and $8$-neighbor lattices. The transition points as well as the
styles of spatial patterns can be explained by local stability
analysis. In sharp contrast to previously reported results,
cooperation is promoted by the spatial structure in the MBSG.
Interestingly, we found that the frequency of cooperation of the
MBSG on a scale-free network peaks at a specific value of the
payoff parameter. This phenomenon indicates that properly
encouraging selfish behaviors can optimally enhance the
cooperation. The memory effects of individuals are discussed in
detail and some non-monotonous phenomena are observed on both
lattices and scale-free networks. Our work may shed some new light
on the study of evolutionary games over networks.
\end{abstract}

\maketitle

\section{Introduction}
Evolutionary game theory has been considered an important approach
to characterizing and understanding the emergence of cooperative
behavior in systems consisting of selfish individuals
\cite{cooperation}. Such systems are ubiquitous in nature, ranging
from biological to economic and social systems. Since the
groundwork on repeated games by Axelrod \cite{Axelrod}, the
evolutionary prisoner's dilemma game (PDG) as a general metaphor
for studying the cooperative behavior has drawn much attention
from scientific communities. Due to the difficulties in assessing
proper payoffs, the PDG has some restriction in discussing the
emergence of cooperative behavior. Thus, the proposal of the
snowdrift game (SG) was generated to be an alternative to the PDG.
The SG, equivalent to the hawk-dove game, is also of biological
interest \cite{SG}. However, in these two games, the unstable
cooperative behavior is contrary to the empirical evidence. This
disagreement motivates a number of extensions of the original
games to provide better explanations for the emergence of
cooperation \cite{Axelrod,Nowak1}.

The spatial game, introduced by Nowak and May \cite{Nowak2}, is a
typical extension, which can result in emergence and persistence
of cooperation in the PDG. Motivated by the idea of the spatial
game, many interests have been given to the effects of spatial
structures, such as lattices \cite{lattice} and networks
\cite{network}, on cooperative behavior. In a recent paper
\cite{Hauert}, Hauert and Doebeli found that compared to the PDG,
cooperation is inhibited by the spatial structure. The surprising
finding is in sharp contrast to one's intuition, since in
comparison with the PDG, the SG favors cooperation. More recently,
Santos and Pacheco \cite{Santos} discovered that scale-free
networks provide a unified framework for the emergence of
cooperation. Besides, Szab\'o et al. \cite{Szabo} presented a
stochastic evolutionary rule to capture the bounded rationality of
individuals for better characterizing the dynamics of games in
real systems.

Among the previous work, the effects of individuals' memory have
not received much attention in the study of evolutionary games on
networks. We argue that individuals usually make decisions based
on the knowledge of past records in nature and society, and the
historical memory would play a key role in an evolutionary game.
Therefore, in the present work, we propose a memory-based
snowdrift game (MBSG), in which players update their strategies
based on their past experience. Our work is partially inspired by
Challet and Zhang \cite{zhang}, who presented a so-called minority
game, in which agents make decisions exclusively according to the
common information stored in their memories. It is found that
finite memories of agents have crucial effects on the dynamics of
the minority game \cite{MG}. We focus on the evolutionary SG for
its general representation of many social and biological
scenarios. The MBSG on different network structures, including
lattices and scale-free networks, is studied. Transitions of
spatial patterns with relevant sudden decreases of the frequencies
of cooperation are observed in lattices. Local stability analyses
are provided for explaining such phenomena. In a scale-free
network, cooperation level peaks at a specific value of payoff
parameter, which is different from previously reported results.
For both lattices and scale-free networks, we found that memory
effects play different roles on the frequency of cooperation for
distinct ranges of the payoff parameter.

\section{the model}
We first briefly describe the original SG model. Imagine that two
cars are trapped on either side of a huge snowdrift. Both drivers
can either get out of the car to shovel (cooperate-C) or stay in
the car (defect-D) in any one negotiation. If they both choose C,
then they both gain benefit $b$ of getting back home while sharing
labor $c$ of shovelling, i.e., both get payoff $b-c/2$. If both
drivers choose D, they will still be trapped by the snowdrift and
get nothing. If one shovels (C) while the other one stays in the
car (D), then they both can get home but the defector pays no
labor cost and gets a perfect payoff $b$, while the cooperator
gets $b-c$. Without losing generality, $b-c/2$ is usually set to
be $1$ so that the evolutionary behavior of the SG can be
investigated with a single parameter, $r=c/2=c/(2b-c)$. Thus, one
has a rescaled payoff matrix
\begin {eqnarray}
\begin {array}{c c c c}
&{\bf C} & {\bf D}\\ {\bf C} & 1 & 1-r \\
 {\bf D} & 1+r & 0  \nonumber
\end {array}
\end {eqnarray}
where $0<r<1$. Though, compared with the PDG, the payoff rank of
the SG favors the emergence of cooperation, cooperation is still
unstable, which results from the highest payoff of defectors.

Here, we introduce the rules of the evolutionary MBSG. Consider
that $N$ players are placed on the nodes of a certain network. In
every round, all pairs of connected players play the game
simultaneously. The total payoff of each player is the sum over
all its encounters. After a round is over, each player will have
the strategy information (C or D) of its neighbors. Subsequently,
each player knows its best strategy in that round by means of
self-questioning, i.e., each player adopts its anti-strategy to
play a virtual game with all its neighbors, and calculates the
virtual total payoff. Comparing the virtual payoff with the actual
payoff, each player can get its optimal strategy corresponding to
the highest payoff and then record it into its memory. Taking into
account the bounded rationality of players, we assume that players
are quite limited in their analyzing power and can only retain the
last $M$ bits of the past strategy information. At the start of
the next generation, the probability of making a decision
(choosing C or D) for each player depends on the ratio of the
numbers of C and D stored in its memory, i.e.,
$P_C=\frac{N_C}{N_C+N_D}=\frac{N_C}{M}$ and $P_D=1-P_C$, where
$N_C$ and $N_D$ are the numbers of C and D, respectively. Then,
all players update their memories simultaneously. Repeat the above
process and the system evolves.

\begin{figure}
\scalebox{0.80}[0.80]{\includegraphics{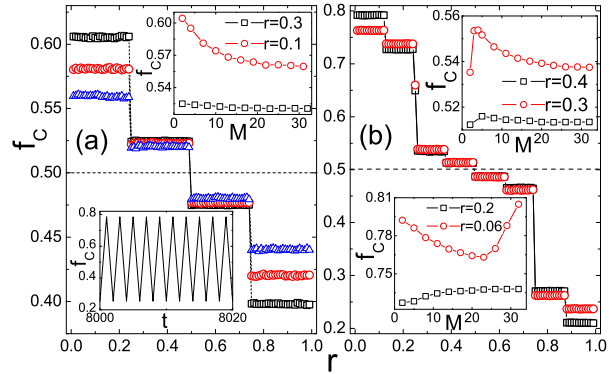}}
\caption{\label{fig:epsart} (Color online). The frequency of
cooperation $f_C$ as a function of the payoff parameter $r$ for
two-dimensional (a) $4$-neighbor and (b) $8$-neighbor lattices,
respectively. ($\bigtriangleup$), ($\bigcirc$) and ($\square$) are
for $M=2$, $7$ and $30$, respectively. Each data point is obtained
by averaging over $40$ different initial states and $f_C$ for each
simulation is obtained by averaging from MC time step $t=5000$ to
$t=10000$, where the system has reached a steady state. The top
inset of (a) is $f_C$ as a function of memory length $M$ for $2$
different cooperation levels. The bottom inset of (a) is a time
series of $f_C$ for $r=0.4$ in the case of $M=1$. Since for $M=1$,
$f_C$ as a function of $t$ displays a big oscillation, we do not
compute the $f_C$ over a period of MC time steps. The inset of (b)
is $f_C$ depending on $M$ for 4 cooperation levels in the range of
$0<r<0.5$. The network size is $N=10000$.}
\end{figure}

\section{MBSG on lattices}
The key quantity for characterizing the cooperative behavior is
the frequency of cooperation, $f_C$, which is defined as the
fraction of C in the whole population. $f_C$ can be obtained by
counting the number of cooperators in the whole population after
the system reaches a steady state, at which the number of
cooperators shows slight fluctuations around an average value.
Hence, $f_C$ is the ratio of the cooperator number and the total
number of individuals $N$. One can easily see that $f_C$ ranges
from $0$ to $1$, where $0$ and $1$ correspond to cases of no
cooperators and entire cooperator state. Firstly, we investigate
the MBSG on two-dimensional square lattices of four and eight
neighbors with periodic boundary conditions. Simulations are
carried out for a population of $N=10000$ individuals located on
nodes. Initially, the strategies of C and D are uniformly
distributed among all players. The memory information of each
player is randomly assigned, and we have checked that this
assignment has no contributions to the stable behavior of the
system. Each data point is obtained by averaging over $40$
different initial states. Figures 1 (a) and (b) show $f_C$ as a
function of the parameter $r$ on the lattices of four and eight
neighbors, respectively. In these two figures, four common
features should be noted: (i) $f_C$ has a step structure, and the
number of steps corresponds to the number of neighbors on the
lattice, i.e., $4$ steps for the $4$-neighbor lattice and $8$
steps for the $8$-neighbor lattice; (ii) the two figures have
$180^{\circ}$-rotational symmetry about the point ($0.5$, $0.5$);
(iii) the memory length $M$ has no influences on the dividing
point $r_c$ between any two cooperation levels, but has strong
effects on the value of $f_C$ in each level; (iv) for a large
payoff parameter $r$, the system still behaves in a high
cooperation level, contrary to the results reported in
\cite{Hauert}. It indicates that although selfish individuals make
decisions based on the best choices stored in their memories to
maximize their own benefits, the cooperative behavior can emerge
in the population in spite of the highest payoff of D.

\begin{figure}
\scalebox{0.91}[0.90]{\includegraphics{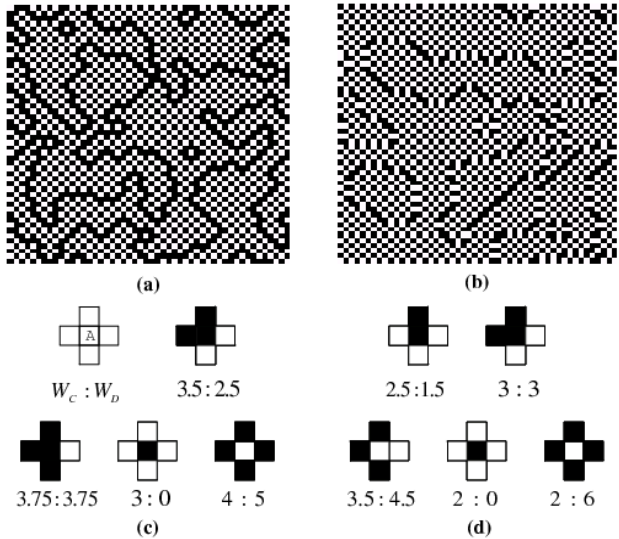}}
\caption{\label{fig:epsart} Typical spatial patterns in two distinct
payoff parameter ranges: (a) $0<r<0.25$, (b) $0.25<r<0.5$. The C is
in black and the D is in white. A $50\times 50$ portion of the full
$100\times 100$ lattice with $4$ neighbors is illustrated. (c) and
(d) are the relevant stable local patterns of (a) and (b). $W_C$ and
$W_D$ are the payoffs of the center individual A by choosing C and D
with fixing strategies of neighbors for each local pattern. $r=0.25$
in (c) and $0.5$ in (d).}
\end{figure}

The effects of memory length $M$ on $f_C$ in the $4$-neighbor
lattice are shown in the insets of Fig. 1. Since $f_C$ is
independent of $r$ within each cooperation level, we simply choose
a value of $r$ in each level to investigate the influence of $M$
on $f_C$. Moreover, due to the inverse symmetry of $f_C$ about the
point ($0.5, 0.5$), we concentrate on the range of $0<r<0.5$. The
top inset of Fig. 1 (a) reports $f_C$ as a function of $M$ for the
ranges of $0<r<0.25$ and $0.25<r<0.5$. One can find that $f_C$ is
a monotonous function of $M$ for both levels and the decreasing
velocity of $f_C$ in the $1$st level is faster than that in the
$2$nd one. In contrast, in the $8$-neighbor lattice, $f_C$
exhibits some non-monotonous behaviors as $M$ increases. As shown
in the bottom inset of Fig. 1 (b), there exists a minimum $f_C$ in
the $1$st level corresponding to $M=23$, and $f_C$ is an
increasing function of $M$ in the $2$nd level. A maximum value of
$f_C$ exists in the $3$rd and $4$th levels when $M$ is chosen to
be $5$, as shown in the top inset of Fig. 1 (b). Thus, memory
length $M$ plays a very complex role in $f_C$ reflected by the
remarkably different behaviors in $4$ cooperation levels. It is
worth to point out that in case of $M=1$, the evolutionary
behavior of the system sharply differs from that of $M>1$, since
each individual will definitely adopt the exclusive strategy
stored in its memory to play the game at the next time step. A
typical example with $M=1$ for two types of lattices is shown in
the bottom inset of Fig. 1 (a). A big oscillation of $f_C$ is
observed. The unstable behavior will be explained in terms of the
evolution of spatial patterns later.

We give a heuristic analysis of local stability for the dividing
points $r_c$ of different levels. At each critical point $r_c$
between any two levels, the payoff of an individual with strategy
C should equal that of the individual with D. We assume the number
of C neighbors of a given node to be $m$, thus in the $K$-neighbor
lattice, the quantity of defector neighbor is $K-m$. Accordingly,
we get the local stability equation: $m+(K-m)(1-r_c)=(1+r_c)m$,
where the left side is the payoff of the given individual with C,
and the right side is the payoff of the individual with D. This
equation results in $r_c=(K-m)/K$. Considering all of the possible
values of $m$ in the $4$-neighbor lattice, the values of $r_c$ are
$0.25$, $0.5$ and $0.75$, respectively. Similarly, the dividing
points of the $8$-neighbor lattice are obtained as $1/8,
2/8,\cdots, 7/8$. As shown in Figs. 1 (a) and (b), the simulation
results are in good accordance with the analytical predictions.
Moreover, it should be noted that there exists a sharp decrease of
$f_C$ at $r_c$, which implies the sudden transformation of the
evolutionary pattern of the system.

\begin{figure}
\scalebox{0.9}[1.0]{\includegraphics{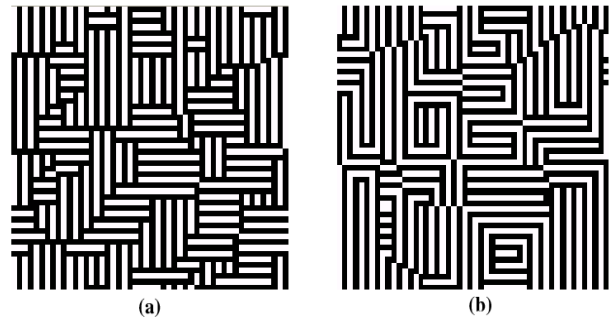}}
\caption{\label{fig:epsart} (Color online). Typical spatial patterns
in two distinct payoff parameter ranges: (a) $0.25<r<0.375$; (b)
$0.375<r<0.5$. The color coding is the same as Fig. 2. A $50\times
50$ portion of the full $100\times 100$ lattice with $8$ neighbors
is illustrated.}
\end{figure}

To gain some intuitionistic insights into the evolution of the
system, we investigate the spatial patterns for different $r$ on
lattices. Figure 2 illustrates typical patterns of two cooperation
levels on the $4$-neighbor lattice. The patterns are statistically
static, independent of initial states. Figure 2 (a), for
$0<r<0.25$, is a typical spatial pattern of `C lines' against a
background of `chessboard' form, i.e., a site is surrounded by
anti-strategy neighbors. Figure 2 (b) is for the range of
$0.25<r<0.5$. In contrast to Fig. 2 (a), `C lines' are broken in
some places by D sites, and some flower-like local patterns are
observed. The patterns in the ranges of $0.5<r<0.75$ and
$0.75<r<1$ are the patterns of Figs. 2 (b) and (a) with C and D
site exchanged, respectively, which are not shown here. Therefore,
there exist four kinds of spatial patterns with typical features
corresponding to four levels of $f_C$. The pattern formation can
be explained in terms of steady local patterns. In Fig. 2 (c), we
show the steady local patterns existing in the $1$st cooperation
level. From the payoff ratio by choosing C and D of individual A,
i.e., $W_C: W_D$, the $3$rd local pattern is the most stable one
with the highest payoff ratio. In parallel, the $4$th local
pattern is the counterpart of the $3$nd one, so that it is also
very stable. Hence, the pattern in Fig. 2 (a) has a
chessboard-like background together with C lines composed of the
$1$st and $2$nd local patterns. Similarly, the chessboard-like
background in Fig. 2 (b) is also attributed to the strongest
stability of the $4$th and $5$th local patterns, and the
probability of the occurrence of other local patterns is
correlated with their payoff ratios. Whereafter, we study the
spatial patterns on the $8$-neighbor lattice. In Fig. 3, we
figured out that each cooperation level exhibits a unique pattern
and the difference between the patterns of $r<0.5$ and $r>0.5$ is
the exchange of C and D sites. For the $1$st and $2$nd levels, D
sites take the minority and submerge into the ocean of C sites.
While in the $3$rd and $4$th levels, interesting patterns emerge.
As shown in Fig. 3 (a), D sites form zonary shapes, surrounded by
C lines. Figure 3 (b) is for the range of $0.375<r<0.5$. The
pattern shows a shape of labyrinth, and the fraction of C sites is
slightly larger than that of D sites. The pattern style can also
be explained by the stability of local patterns as that in the
$4$-neighbor lattice.

\begin{figure}
\scalebox{0.95}[1.0]{\includegraphics{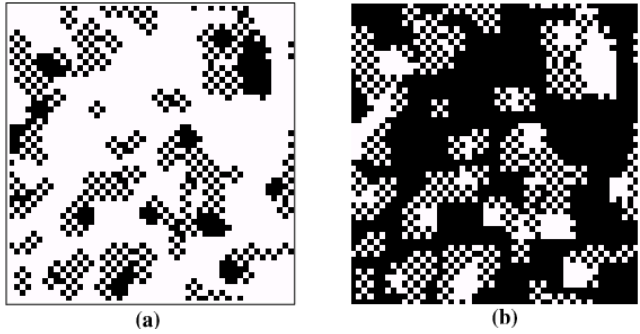}}
\caption{\label{fig:epsart} (Color online). Typical patterns for the
time step $t=8001$ and $t=8002$ in the case of memory length $M=1$,
$r=0.4$. The C is in black and the D is in white. A $50\times 50$
portion of the full $100\times 100$ lattice with $4$ neighbors is
illustrated.}
\end{figure}

We have discussed the static patterns on lattices, next we will
provide a description of patterns in the case of $M=1$, where the
patterns are unstable, reflected by a big oscillation in the inset
of Fig. 1 (a). Two typical patterns for $M=1$ on a $4$-neighbor
lattice are displayed in Fig. 4. One can see that a large fraction
of adjacent defectors (denoted by the white area in Fig. 4 (a))
switch to cooperators together at the next time step (denoted by
the large area in black in Fig. 4 (b)), which contributes to the
big oscillation of $f_C$. The strategy-switch behavior of large
proportional individuals can be easily explained by noting the
fact that individuals will update their strategies by adopting the
exclusive strategy in their memories ($M=1$). Moreover,
individuals record the strategy on the basis of their neighbors'
strategy at the last time step. In the strategy-switch area,
individuals have identical strategies at each time step.
Therefore, at the next step, each individual should choose the
anti-strategy of its neighbors to gain more payoffs since each one
only records its last step's history. Once the drastic strategy
switch occurs, it will maintain forever.

In addition, we should briefly introduce a recent work of Sysi-Aho
et al. \cite{myopic}, which is correlated with the present MBSG
model. In Ref. \cite{myopic}, the authors proposed a spatial
snowdrift game played by myopic agents. In such model, lattices
are used and each individual can adopt its current anti-strategy
at the next time according to its neighbors' strategies with a
probability $p$. Similar spatial patterns are observed for
$8$-neighbor lattices, as well as the step structure of $f_C$
depending on $r$. However, we note that $p$ in this model nearly
has no effect on the cooperative behavior, while in our model the
memory length $M$ plays different roles in each cooperation level.
Furthermore, in the case of no memory length, i.e., $M=1$, our
model doesn't recover the spatial snowdrift game with myopic
agents, confirmed by the big oscillation of $f_C$ in the inset of
Fig. 1 (a).

\begin{figure}
\scalebox{1.08}[1.08]{\includegraphics{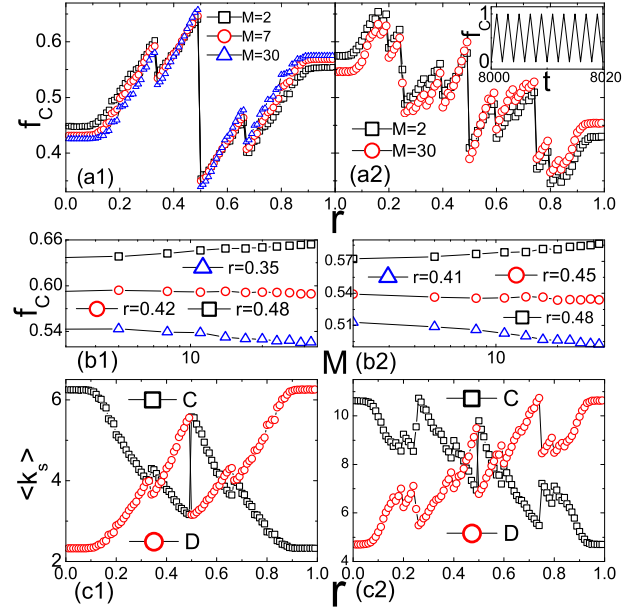}}
\caption{\label{fig:epsart} (Color online). $f_C$ as a function of
$r$ in BA networks with (a1) average degree $\langle k\rangle=4$
and (a2) $\langle k\rangle=8$ for different $M$. A time series of
$f_C$ for $M=1$ is shown in the inset of (a2). (b1) and (b2) are
$f_C$ as a function of $M$ in the case of $\langle k\rangle=4$ and
$\langle k\rangle=8$ for a special range of $r$. (c1) and (c2) are
average degrees $\langle k_s\rangle$ of C and D players depending
on $r$ in the case of $M=7$ for $\langle k\rangle=4$ and $\langle
k\rangle=8$, respectively. The network size is $10000$. Each data
point is obtained by averaging over $30$ different network
realizations with $20$ different initial state of each
realization. $f_C$ for each simulation is obtained by averaging
from MC time step $t=5000$ to $t=10000$, where the system has
reached a steady state.}
\end{figure}

\section{MBSG on scale-free networks}
Going beyond two-dimensional lattices, we also investigate the
MBSG on scale-free (SF) networks, since such structural property
is ubiquitous in natural and social systems. Figure 5 shows the
simulation results on the Barab\'asi-Albert networks \cite{BA},
which are constructed by the preferential attachment mechanism.
Each data point is obtained by averaging over $30$ different
network realizations with $20$ different initial states of each
realization. Figures 5 (a1) and (a2) display $f_C$ depending on
$r$ on BA networks in the cases of average degree $\langle
k\rangle=4$ and $\langle k\rangle=8$ for different memory lengths
$M$. There are some common features in these two figures: (i) in
sharp, contrast to the cases on lattices, $f_C$ is a
non-monotonous function of $r$ with a peak at a specific value of
$r$. This interesting phenomenon indicates that properly
encouraging selfish behaviors can optimally enhance the
cooperation on SF networks; (ii) it is the same as the cases on
lattices that the continuity of $f_C$ is broken by some sudden
decreases. The number of continuous sections corresponds to the
average degree $\langle k\rangle$; (iii) two figures have a
$180^{\circ}$-rotational symmetry about the point ($0.5$, $0.5$);
(iv) the memory length $M$ does not influence the values of $r$,
at which sudden decreases occur, as well as the trend of $f_C$,
but affects the values of $f_C$ in each continuous section. Then,
we investigate the effect of $M$ on $f_C$ in detail. Due to the
inverse symmetry of $f_C$ about point $(0.5, 0.5)$, our study
focus on the range of $0<r<0.5$. We found that in both SF
networks, there exists a unique continuous section, in which $M$
plays different roles in $f_C$. For the case of $\langle
k\rangle=4$, the special range is from $r=0.34$ to $0.49$, as
shown in Fig. 5 (a1). In this region $f_C$ as a function of $M$ is
displayed in Fig. 5 (b1). One can find that for $r=0.42$, $f_C$ is
independent of $M$. For $0.34<r<0.42$, $f_C$ is a decreasing
function of $M$; while for $0.42<r<0.49$, $f_C$ becomes an
increasing function of $M$. Similar phenomena are observed in the
SF network with $\langle k\rangle=8$, as exhibited in Fig. 5 (b2).
$r=0.45$ is the dividing point, and for $r<0.45$ and $r>0.45$,
$f_C$ shows decreasing and increasing behaviors respectively as
$M$ increases. In the case of $M=1$, the system has big
oscillations as shown in the inset of Fig. 5 (a2). Similar to the
cases on lattices, the behavior of large proportion of
individuals' strategy switches that induces the big oscillation of
$f_C$ in the SF network.

\begin{figure}
\scalebox{0.78}[0.78]{\includegraphics{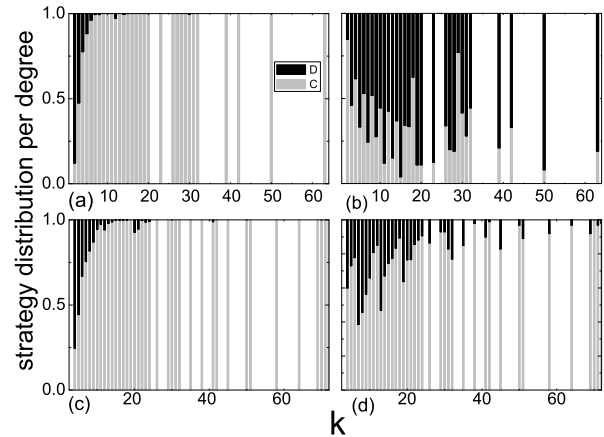}}
\caption{\label{fig:epsart} (Color online). Distributions of
strategies in BA networks. Cooperators and defectors are denoted
by gray bars and black bars, respectively. Each bar adds up to a
total fraction of $1$ per degree, the gray and black fractions
being directly proportional to the relative percentage of the
respective strategy for each degree of connectivity $k$. (a) is
for the case of $\langle k\rangle=4$ with $r=0.1$ and (b) is for
the case of $\langle k\rangle=4$ with $r=0.49$, at which $f_C$
peaks. (c) shows the case of $\langle k\rangle=8$ with $r=0.05$
and (d) displays the case of $\langle k\rangle=8$ with $r=0.16$,
which corresponds to the maximum value of $f_C$. All the
simulations are obtained for network size $N=1000$ in order to
make figures clearly visible.}
\end{figure}

In order to give an explanation for the non-monotonous behaviors
reported in Figs. 5 (a1) and (a2), we study the average degree
$\langle k_s\rangle$ of cooperators and defectors depending on
$r$. In Figs. 5 (c1) and (c2), $\langle k_s\rangle$ of D vs $r$
shows almost the same trend as that of $f_C$ in Figs. 5 (a1) and
(a2), also the same sudden decreasing points at specific values of
$r$. When $r$ is augmented from $0$, large-degree nodes are
gradually occupied by D, reflected by the enhancement of D's
$\langle k_s\rangle$. The detailed description of the occupation
of nodes with given degree can be seen in Fig. 6. One can clearly
find that on the $4$-neighbor lattice, in the case of low value of
$f_C$ (Fig. 6 (a)), almost all high degree nodes are occupied by
cooperators and most low degree nodes are occupied by defectors;
while at the peak value of $f_C$ (Fig. 6 (b)), cooperators on most
high degree nodes are replaced by defectors and on low degree
nodes cooperators take the majority. Similarly, as $f_C$ increases
in the $8$-neighbor lattices, defectors gradually occupy those
high degree nodes, together with most very low degree nodes taken
by cooperators (Fig. 6 (c) and (d)). Moreover, note that in SF
networks, large-degree nodes take the minority and most neighbors
of small-degree nodes are those large-degree ones, so that when
more and more large-degree nodes are taken by D, more and more
small-degree nodes have to choose C to gain payoff $1-r$ from each
D neighbor. Thus, it is the passive decision making of
small-degree nodes which take the majority in the whole
populations that leads to the increase of $f_C$. However, for very
large $r$, poor benefit of C results in the reduction of $f_C$.
Therefore, $f_C$ peaks at a specific value of $r$ on SF networks.
In addition, it is worthwhile to note that in the case of high
$f_C$, the occupation of large degree nodes in the MBSG on SF
networks is different from recently reported results in Ref.
\cite{occupy}. The authors found that all (few) high degree nodes
are occupied by cooperators, whereas defectors only manage to
survive on nodes of moderate degree. While in our work, defectors
take over almost all high degree nodes, which induces a high level
of cooperation.

\section{conclusion}
In conclusion, we have studied the memory-based snowdrift game on
networks, including lattices and scale-free networks. Transitions
of spatial patterns are observed on lattices, together with the
step structure of the frequency of cooperation versus the payoff
parameter. The memory length of individuals plays different roles
at each cooperation level. In particular, non-monotonous behavior
are found on SF networks, which can be explained by the study of
the occupation of nodes with give degree. Interestingly, in
contrast to previously reported results, in the memory-based
snowdrift game, the fact of high degree nodes taken over by
defectors leads to a high cooperation level on SF networks.
Furthermore, similar to the cases on lattices, the average degrees
of SF networks is still a significant structural property for
determining cooperative behavior. The memory effect on cooperative
behavior investigated in our work may draw some attention from
scientific communities in the study of evolutionary games.

This work is funded by the Hong Kong Research Grants Council under
the CERG Grant CityU No. 1114/05E and by the National Natural
Foundation of China under Grant Nos. 704710333, 10472116,
10532060, 70571074, 10547004, and 10635040.

\end{document}